\documentclass[preprint,12pt]{elsarticle}

\usepackage{amssymb}

\usepackage{dcolumn}

\usepackage{graphicx}

\biboptions{sort&compress}

\begin{document}

\begin{frontmatter}

\title{Is space-time symmetry a suitable generalization of parity-time symmetry?}

\author{Paolo Amore\dag} \ead{paolo.amore@gmail.com}

\author{Francisco M.
Fern\'andez\ddag}\ead{fernande@quimica.unlp.edu.ar}

\author{Javier Garcia\ddag}

\address{\dag\ Facultad de Ciencias, CUICBAS,
Universidad de Colima, Bernal D\'{\i}az del Castillo 340, Colima,
Colima, Mexico}

\address{\ddag\ INIFTA (UNLP, CCT La Plata-CONICET), Divisi\'{o}n Qu\'{i}mica Te\'{o}rica,
Diag. 113 y 64 (S/N), Sucursal 4, Casilla de Correo 16, 1900 La
Plata, Argentina}

\begin{abstract}
We discuss space-time symmetric Hamiltonian operators of the form $%
H=H_{0}+igH^{\prime }$, where $H_{0}$ is Hermitian and $g$ real. $H_{0}$ is
invariant under the unitary operations of a point group $G$ while $H^{\prime
}$ is invariant under transformation by elements of a subgroup $G^{\prime }$
of $G$. If $G$ exhibits irreducible representations of dimension greater
than unity, then it is possible that $H$ has complex eigenvalues for
sufficiently small nonzero values of $g$. In the particular case that $H$ is
parity-time symmetric then it appears to exhibit real eigenvalues for all $%
0<g<g_{c}$, where $g_{c}$ is the exceptional point closest to the origin.
Point-group symmetry and perturbation theory enable one to predict whether $%
H $ may exhibit real or complex eigenvalues for $g>0$. We illustrate the
main theoretical results and conclusions of this paper by means of two- and
three-dimensional Hamiltonians exhibiting a variety of different point-group
symmetries.
\end{abstract}


\end{frontmatter}

\section{Introduction}

\label{sec:intro}

In the last years there has been great interest in the properties of
PT-symmetric multidimensional oscillators\cite
{BDMS01,NA02,N02,N05,BTZ06,W09,CIN10,BW12,HV13}. Among them we mention the
complex versions of the Barbanis\cite{BDMS01,NA02,N05,BTZ06,W09,BW12,HV13}
and H\'{e}non-Heiles\cite{BDMS01,W09} Hamiltonians. Several methods have
been applied to the calculation of their spectra: the diagonalization method%
\cite{BDMS01,NA02,N02,N05,W09,BW12}, perturbation theory\cite
{BDMS01,N02,N05,W09}, classical and semiclassical approaches\cite
{BDMS01,NA02}, among others\cite{W09,HV13}. Typically, those models depend
on a potential parameter $g$ so that the Hamiltonian is Hermitian when $g=0$
and non-Hermitian when $g\neq 0$. Bender and Weir\cite{BW12} conjectured
that the models studied so far may exhibit PT phase transitions so that
their spectra are entirely real for sufficiently small but nonzero values of
$|g|$. Such phase transitions appear to be a high-energy phenomenon and take
place at exceptional points\cite{HS90,H00,HH01,H04}. More precisely: as $g$
increases two real eigenvalues approach each other, coalesce at an
exceptional point $g_{c}$ and become a pair of complex conjugate numbers for
$g>g_{c}$. The PT phase transition takes place at the smallest $g_{c}$.

Multidimensional oscillators exhibit point-group symmetry (PGS)\cite
{PE81a,PE81b}. Klaiman and Cederbaum\cite{KC08} were the first to apply PGS
to non-Hermitian Hamiltonians of the form $H_{0}+i\lambda W$ to predict the
symmetry of the eigenfunctions associated to the eigenvalues that coalesce
at the exceptional points. These authors proposed an interesting approach to
study such points in terms of an effective Hermitian operator built from the
Hermitian $H_{0}$ and non-Hermitian $W$ parts of the original Hamiltonian
operator. They also coined the term space-time symmetry that refers to a
class of antiunitary symmetries that contain the PT symmetry as a particular
case. The analysis of Klaiman and Cederbaum\cite{KC08} was restricted to
Abelian point groups that exhibit only one-dimensional irreducible
representations (irreps).

The main interest in the study of PT-symmetric oscillators has been to
enlarge the class of non-Hermitian Hamiltonians that exhibit real spectra,
at least for some values of the potential parameter $g$ (or $\lambda $). In
such cases PT symmetry (or more generally ST symmetry) is broken at the
exceptional points $g_{c}$ already mentioned above which can be efficiently
calculated as critical parameters by means of the diagonalization method\cite
{FG13a}. The PT phase transition is determined by the smallest $|g_{c}|$.

By means of PGS Fern\'{a}ndez and Garcia\cite{FG14,FG14b} found some
examples of ST-symmetric multidimensional oscillators that exhibit complex
eigenvalues for $g>0$ so that the phase transition appears to take place at
the trivial Hermitian limit $g=0$. Their results suggest that the more
general ST symmetry is not as robust as the PT one and contradict some of
the conjectures put forward by Klaiman and Cederbaum\cite{KC08} based on
PGS. In this paper we discuss this point in more detail, improve and extend
the results and conclusions of those two papers, and look for more
ST-symmetric models with broken ST symmetry for all values of the parameter $%
g$ that measures the strength of the non-Hermitian part. In Section~\ref
{sec:PT} we argue that perturbation theory is suitable to guess whether ST
symmetry is broken at the Hermitian limit $g=0$ or at an exceptional point $%
g=g_{c}>0$. In Section~\ref{sec:symmetry} we outline the main ideas of
unitary and antiunitary symmetry in a way that improves the discussion in
the earlier papers\cite{FG13a,FG14}. In Section~\ref{sec:diagonalization} we
summarize some well known results about the application of the
diagonalization method with symmetry-adapted basis sets. In sections \ref
{sec:2D_models} and \ref{sec:3D_models} we illustrate the main ideas of
sections \ref{sec:PT}, \ref{sec:symmetry} and \ref{sec:diagonalization} by
means of suitably chosen examples in two and three dimensions, respectively.
Finally, in Section~\ref{sec:conclusions} we summarize the main results and
draw conclusions.

\section{Perturbation theory}

\label{sec:PT}

Consider a Hamiltonian operator of the form
\begin{equation}
H(\lambda )=H_{0}+\lambda H^{\prime },  \label{eq:H_gen}
\end{equation}
where $UH^{\prime }U^{-1}=-H^{\prime }$ for some unitary transformation $U$ (%
$U^{-1}=U^{\dagger }$). If $H_{0}$ is invariant under $U$ ($%
UH_{0}U^{-1}=H_{0}$) then $UH(\lambda )U^{-1}=H(-\lambda )$.

It follows from $H(\lambda )\psi _{n}(\lambda ,\mathbf{r})=E_{n}(\lambda
)\psi _{n}(\lambda ,\mathbf{r})$ and the expression above that $UH(\lambda
)\psi _{n}(\lambda ,\mathbf{r})=H(-\lambda )U\psi _{n}(\lambda ,\mathbf{r}%
)=E_{n}(\lambda )U\psi _{n}(\lambda ,\mathbf{r})$. We appreciate that $U\psi
_{n}(\lambda ,\mathbf{r})$ is an eigenfunction $\psi _{m}(-\lambda ,\mathbf{r%
})$ of $H(-\lambda )$ with eigenvalue $E_{m}(-\lambda )=E_{n}(\lambda )$.
Since this equality holds for all $\lambda $ we conclude that $%
E_{n}(0)=E_{m}(0)$. Therefore, if $H_{0}$ does not exhibit degenerate
eigenfunctions then $m=n$, $E_{n}(\lambda )=E_{n}(-\lambda )$, and the
perturbation expansion for this eigenvalue will only exhibit even powers of
the perturbation parameter:
\begin{equation}
E_{n}(\lambda )=\sum_{j=0}^{\infty }E_{n}^{(2j)}\lambda ^{2j}.
\label{eq:En_ptseries}
\end{equation}
When $\lambda =ig$ is imaginary ($g$ real) this last equation suggests that
the eigenvalues of the non-Hermitian operator $H(\lambda )$ may be real for
sufficiently small values of $|g|$. Furthermore, if $T$ is the time-reversal
operator\cite{P65} then $A=TU$ is an antiunitary transformation that leaves
the Hamiltonian $H$ invariant $AHA^{-1}=H$ and we say that it is ST symmetric%
\cite{KC08}. For a detailed discussion of antiunitary operators see the
paper by Wigner\cite{W60}.

The situation may be quite different when $H_{0}$ exhibits degenerate
eigenfunctions
\begin{equation}
H_{0}\psi _{n,i}^{(0)}=E_{n}^{(0)}\psi _{n,i}^{(0)},\;n=0,1,\ldots
,\;i=1,2,\ldots ,\nu _{n}.  \label{eq:H0_psi_n,j}
\end{equation}
If there are nonzero matrix elements of the form
\begin{equation}
H_{ij}^{\prime }=\left\langle \psi _{n,i}^{(0)}\right| H^{\prime }\left|
\psi _{n,j}^{(0)}\right\rangle \neq 0,1\leq i,j\leq \nu _{n}  \label{eq:H'ij}
\end{equation}
then some of the perturbation corrections of first order may be nonzero and
the corresponding eigenvalues
\begin{equation}
E_{n,j}=E_{n}^{(0)}+E_{n,j}^{(1)}\lambda +\ldots  \label{eq:Enj_ptseries}
\end{equation}
may be complex, at least for sufficiently small values of $|g|$. In other
words: one expects broken ST symmetry for $g>0$ when $H_{0}$ exhibits
degenerate eigenfunctions with nonzero matrix elements $H_{ij}^{\prime }$.
As we will see below, PGS is most helpful for finding such examples.

\section{Unitary and antiunitary symmetry}

\label{sec:symmetry}

In this paper we consider Hamiltonian operators of the form (\ref{eq:H_gen})
where $\lambda =ig$, $g$ real. We assume that $H_{0}$ is Hermitian and
invariant under the operations of the group $G=\left\{ U_{1},U_{2},\ldots
,U_{m}\right\} $: $U_{i}H_{0}U_{i}^{-1}=H_{0}$ (in this paper we restrict
ourselves to point groups\cite{T64,C90}). If $H^{\prime }$ is invariant
under the operations of a subgroup $G^{\prime }=\left\{ W_{1},W_{2},\ldots
,W_{k}\right\} $ of $G$ ($W_{i}H^{\prime }W_{i}^{-1}=H^{\prime }$) then $H$
is invariant under the operations of the point group $G^{\prime }$.

Suppose that $U_{i}H^{\prime }U_{i}^{-1}=-H^{\prime }$, where $U_{i}\in
G\backslash G^{\prime }$. Then the Hamiltonian exhibits an antiunitary
symmetry (space-time symmetry) given by $\hat{A}_{i}=TU_{i}$; that is to
say, $H$ is invariant under $\hat{A}_{i}$: $\hat{A}_{i}H\hat{A}_{i}^{-1}=H$.
Because of this antiunitary symmetry the eigenvalues of $H$ are either real
or appear in pairs of complex conjugate numbers. In fact, if $\psi $ is an
eigenfunction of $H$ with eigenvalue $E$ and $\hat{A}$ is an antiunitary
symmetry of $H$, then
\begin{equation}
H\hat{A}\psi =\hat{A}\hat{A}^{-1}H\hat{A}\psi =\hat{A}H\psi
=E^{*}\hat{A}\psi . \label{eq:HApsi}
\end{equation}
If $\hat{A}\psi =a\psi $ then $E$ is real and we say that the
space-time symmetry is unbroken. It may also be possible that
$\hat{A}\psi $ is a linear combination of degenerate
eigenfunctions of $H$ with eigenvalue $E$ and we arrive at the
same conclusion\cite{FG13a}. Klaiman and Cederbaum\cite {KC08}
coined the term space-time symmetry to indicate an antiunitary
symmetry $\hat{A}=ST$, where the unitary operator $S$ may be other
than the parity operation $P:(x,y,z)\rightarrow (-x,-y,-z)$.
Obviously, ST symmetry contains PT symmetry as a particular case
($S=P$) and it is understood that in the latter case $P$ belongs
to $G$ but not to $G^{\prime}$.

Klaiman and Cederbaum\cite{KC08} argued that in principle one can
get an entirely real spectrum for a non-Hermitian Hamiltonian $H$
if $H^{\prime }$ is chosen such that it transforms as an irrep of
the point group or subgroup of $H_{0}$. They assumed that the
spectrum of $H_{0}$ is nondegenerate, thus restricting themselves
to Abelian groups with real character tables. This restriction is
crucial if $H^{\prime }$ is to transform as one of the irreps of
the point group of $H_{0}$ since degenerate states belonging to
higher dimensional irreps tend to couple to themselves no matter
what irrep one chooses for $H^{\prime }$. They also stated that if
the non-Abelian point group of $H_{0}$ (in the case of a
degenerate spectrum) has an Abelian subgroup of order larger than
$1$, one can still choose $H^{\prime }$ such that it transforms
under the irreps of the Abelian subgroup and $H$ can still, in
principle, have a completely real spectrum. They also pointed out
that if one wishes to keep only part of the spectrum of $H$ on the
real axis, many more options become available. Fern\'{a}ndez and
Garcia\cite {FG14b} discussed the non-Hermitian model given by a
particle in a square box with the perturbation $H^{\prime }=xy$.
In this case the point group for $H_{0}$ is $C_{4v}$ with the
Abelian subgroup $C_{2v}$ of order greater than $1$. $H^{\prime }$
transforms as the irrep $B_{2}$ of $C_{4v}$ and the irrep $A_{2}$
of $C_{2v}$\cite{T64,C90}. However, the spectrum for this model
does not appear to be entirely real because some of the
eigenvalues are complex for arbitrarily small values of $|g|$.

Because of what we have just discussed, in this paper we are mainly
interested in the case that $H_{0}$ exhibits degenerate eigenfunctions (\ref
{eq:H0_psi_n,j}) and $G$ exhibits one or more irreps of dimension greater
than one. As argued in Section~\ref{sec:PT} if there are nonzero matrix
elements of the form (\ref{eq:H'ij}) then some of the perturbation
corrections of first order are nonzero and the corresponding eigenvalues (%
\ref{eq:Enj_ptseries}) are complex for small values of $|g|$. If $\psi
_{n,j}^{(0)}$ and $H^{\prime }$ belong to the irreps $\Gamma _{n}$ and $%
\Gamma _{H^{\prime }}$, respectively, then the matrix elements $%
H_{ij}^{\prime }$ may be nonzero if the decomposition of the reducible
representation $\Gamma _{n}\otimes \Gamma _{n}\otimes \Gamma _{H^{\prime }}$
contains the totally symmetric irrep\cite{T64,C90}. Since $\psi
_{n,i}^{(0)}\psi _{n,j}^{(0)}$ is invariant under $P$, then $H_{ij}^{\prime
} $ vanishes unless $H^{\prime }$ is also parity invariant $PH^{\prime
}P=H^{\prime }$. Therefore, under the latter condition it is likely that an
ST-symmetric Hamiltonian may exhibit complex eigenvalues for sufficiently
small values of $|g|$. On the other hand, all the PT-symmetric Hamiltonians
studied so far exhibit real eigenvalues for $0\leq g<g_{c}$. This point has
already been discussed in two recent papers\cite{FG14,FG14b}.

In addition to the unitary and antiunitary symmetries outlined above it is
worth considering possible dynamical symmetries. If $O$ is an Hermitian
operator that commutes with $H_{0}$ and $\psi ^{(0)}$ is an eigenfunction of
the latter with eigenvalue $E^{(0)}$ then $O\psi ^{(0)}$ is also
eigenfunction of $H_{0}$ with the same eigenvalue as follows from $%
H_{0}O\psi ^{(0)}=OH_{0}\psi ^{(0)}=E^{(0)}O\psi ^{(0)}$. If, in addition, $%
\psi ^{(0)}$ and $O\psi ^{(0)}$ belong to different irreps of the point
group $G$ for $H_{0}$ then the dimension of some of the eigenspaces of this
operator cannot be explained solely by PGS (see \cite
{M59,LFLA97,LFAL98,HCL13,F13b} and the references therein).

\section{Diagonalization method}

\label{sec:diagonalization}

Throughout this paper we calculate the eigenvalues of the non-Hermitian
operator $H$ by means of three approaches: the Riccati-Pad\'{e} method\cite
{FMT89a, FMT89b}, a collocation method\cite{AF10, AFR11}, and the
straightforward diagonalization method\cite{BDMS01,NA02,N02,N05,W09,BW12}
that consists in obtaining the eigenvalues of a truncated matrix
representation of the Hamiltonian operator in a suitable basis set.
Commonly, one chooses a complete set of orthonormal functions $%
F=\{f_{1},f_{2},\ldots \}$ which we can split into subsets of
symmetry-adapted functions $F^{S}=\{f_{1}^{S},f_{2}^{S},\ldots \}$ for each
irrep $S$\cite{T64,C90}. Instead of diagonalizing and $M\times M$ matrix
representation $\mathbf{H}$ of the Hamiltonian operator in the basis set $F$
we diagonalize $M_{S}\times M_{S}$ matrix representations $\mathbf{H}^{S}$ ($%
M_{S}<M$) of $H$ in each basis set $F^{S}$. This strategy not only enables
us to reduce the dimension of the matrices to be diagonalized but also
facilitates the interpretation of the results\cite{FG14,FG14b}.

Every eigenfunction of $H$ that belongs to the irrep $S$ can be written as a
linear combination of the complete set of functions of the corresponding
symmetry:
\begin{equation}
\psi ^{S}=\sum_{j}c_{j}^{S}f_{j}^{S}.  \label{eq:psi^S_expansion}
\end{equation}
Suppose that $\hat{A}=UT$ is an antiunitary symmetry of $H$ such that the
space transformation $U$ changes the symmetry of the basis set according to
\begin{equation}
Uf_{j}^{S}=\sum_{k}d_{kj}^{S^{\prime }S}f_{k}^{S^{\prime }},
\label{eq:Uf_j^S}
\end{equation}
and that $Tf_{j}^{S}=f_{j}^{S}$. Therefore, $\hat{A}\psi ^{S}=\psi
^{S^{\prime }}$ and $H\hat{A}\psi ^{S}=E^{S^{\prime }}\hat{A}\psi ^{S}$. On
the other hand, Equation (\ref{eq:HApsi}) tells us that $HA\psi ^{S}=\left(
E^{S}\right) ^{*}A\psi ^{S}$ and we conclude that $E^{S^{\prime }}=\left(
E^{S}\right) ^{*}$ under the conditions just stated. We will see some
examples of this result in sections \ref{sec:2D_models} and \ref
{sec:3D_models}.

\section{Two-dimensional models}

\label{sec:2D_models}

In this section we consider some two-dimensional examples of the Hamiltonian
(\ref{eq:H_gen}). In order to discuss and illustrate their main ideas
Klaiman and Cederbaum\cite{KC08} chose $H_{0}=\frac{1}{2}\left(
p_{x}^{2}+p_{y}^{2}\right) +\alpha _{x}x^{4}+\alpha _{y}y^{4}$. When $\alpha
_{x}\neq \alpha _{y}$ the point group $G$ for $H_{0}$ is $C_{2v}$ (they
chose $D_{2h}^{2D}$) with only one-dimensional irreps and the numerical
results suggest that the eigenvalues are real for $0<g<g_{c}$, where $g_{c}$
is the exceptional point closest to the origin. In this section we consider
closely related models with different PGS.

The first set of examples that we discuss in what follows is based on the
Hermitian part
\begin{equation}
H_{0}=p_{x}^{2}+p_{y}^{2}+x^{4}+y^{4},  \label{eq:H0_x4_y4}
\end{equation}
which is invariant under the operations $%
\{E,C_{4},C_{4}^{2}=C_{2},C_{4}^{3},\sigma _{v},\sigma _{v}^{\prime },\sigma
_{d},\sigma _{d}^{\prime }\}$ of the symmetry point group $C_{4v}$ shown in
Table~\ref{tab:c4v}. If $\phi _{n}(q)$ is an eigenfunction of $%
p_{q}^{2}+q^{4}$ with eigenvalue $\epsilon _{n}$ then $\varphi
_{m\,n}(x,y)=\phi _{m}(x)\phi _{n}(y)$ is eigenfunction of $H_{0}$ with
eigenvalue $E_{mn}^{(0)}=\epsilon _{m}+\epsilon _{n}$. Linear combinations
of these eigenfunctions are bases for the irreps of the point group $C_{4v}$
according to the following scheme:
\begin{equation}
\begin{array}{ll}
\varphi _{2m\,2m} & A_{1} \\
\varphi _{2m+1\,2m+1} & B_{2} \\
\varphi _{2m\,2n}^{+} & A_{1} \\
\varphi _{2m\,2n}^{-} & B_{1} \\
\varphi _{2m+1\,2n+1}^{+} & B_{2} \\
\varphi _{2m+1\,2n+1}^{-} & A_{2} \\
\left\{ \varphi _{2m\,2n+1},\varphi _{2n+1\,2m}\right\} & E
\end{array}
,  \label{eq:C4v_bases}
\end{equation}
where
\begin{equation}
\varphi _{m\,n}^{\pm }=\frac{1}{\sqrt{2}}\left( \varphi _{m\,n}\pm \varphi
_{n\,m}\right) ,\;m\neq n.
\end{equation}
According to Equation (\ref{eq:C4v_bases}) we expect one-dimensional
eigenspaces of symmetry $A_{1},A_{2},B_{1},B_{2}$ and two-dimensional ones
of symmetry $E$. This is the degeneracy predicted by the geometrical
symmetry of the Hamiltonian operator.

The Hermitian operator
\begin{equation}
O=p_{x}^{2}+x^{4}-p_{y}^{2}-y^{4},
\end{equation}
commutes with $H_{0}$ and connects functions of different symmetry as
follows from
\begin{equation}
O\varphi _{m\,n}^{\pm }=\left( \epsilon _{m}-\epsilon _{n}\right) \varphi
_{m\,n}^{\mp }.
\end{equation}
Since $O$ belongs to the irrep $B_{1}$, $B_{1}\otimes B_{2}=A_{2}$ and $%
B_{1}\otimes B_{1}=A_{1}$, then some functions of symmetry $A_{1}(A_{2})$
are degenerate with functions of symmetry $B_{1}(B_{2})$. Similar dynamical
symmetries for simpler, exactly solvable, two-dimensional models have been
discussed elsewhere\cite{LFLA97, LFAL98}.

The first eigenvalues of the Hermitian Hamiltonian (\ref{eq:H0_x4_y4})
calculated by means of the Riccati-Pad\'{e} method\cite{FMT89a,FMT89b} shown
in Table~\ref{tab:E0x4y4} illustrate the two types of degeneracy
(geometrical and dynamical) just discussed.

If we add the perturbation $H^{\prime }=xy$ then the suitable point group $%
G^{\prime }$ results to be $C_{2v}$ that we modify in order to make it
compatible with the $C_{4v}$ for $H_{0}$. The corresponding modified
character table is shown in Table~\ref{tab:c2v}\ (compare it with the one in
the standard textbooks\cite{T64,C90}). The reflection operators in the $%
C_{4v} $ point group are defined as $\sigma _{v}:(x,y)\rightarrow (-x,y)$, $%
\sigma _{v}^{^{\prime }}:(x,y)\rightarrow (x,-y)$, $\sigma
_{d}:(x,y)\rightarrow (y,x)$ and $\sigma _{d}^{\prime }:(x,y)\rightarrow
(-y,-x)$. Therefore, the antiunitary symmetries $\hat{A}_{1}=T\sigma _{v}$
and $\hat{A}_{2}=T\sigma _{v}^{\prime }$, which satisfy $\hat{A}_{j}^{2}=1$,
leave $H$ invariant: $\hat{A}_{j}H\hat{A}_{j}=H$, $j=1,2$. In this example
of ST symmetry the rotation operation $C_{2}:(x,y)\rightarrow (-x,-y)$ plays
the role of the parity one and leaves the perturbation invariant $%
C_{2}H^{\prime }C_{2}=H^{\prime }$.

It is worth noting that we use the symbols $\sigma _{d}$ and $\sigma
_{d}^{\prime }$ instead of the usual $\sigma _{v}$ and $\sigma _{v}^{\prime }
$ for the reflection planes in the modified character table $C_{2v}$ in
Table~\ref{tab:c2v}. The reason is that we have to define the unitary
operations of the point group $C_{2v}$ so that $H^{\prime }=xy$ belongs to
the totally symmetric irrep $A_{1}$. The point group $C_{2v}$ shown in Table~%
\ref{tab:c2v} plays the role of the subgroup $G^{\prime }$ introduced in the
general discussion of Section~\ref{sec:symmetry}. On the other hand, $%
H^{\prime }$ belongs to the irrep $A_{2}$ of the subgroup $C_{2v}$ that we
obtain by choosing the reflection planes $\sigma _{v}$ and $\sigma
_{v}^{\prime }$. It is clear that in this example $H^{\prime }$ belongs to
an irrep of an Abelian subgroup of order greater than $1$ of the point group
for $H_{0}$. Therefore, $H$ should have real eigenvalues according to
Klaiman and Cederbaum\cite{KC08}.

$H^{\prime }=xy$ belongs to the irrep $B_{2}$ of the point group $C_{4v}$.
Since $E\otimes E=A_{1}\oplus A_{2}\oplus B_{1}\oplus B_{2}$ we conclude
that two degenerate eigenfunctions of $H_{0}$ that are basis for the irrep $%
E $ will lead to nonzero perturbation corrections of first order and,
according to the discussion in Section~\ref{sec:PT}, to complex eigenvalues.
More precisely, the perturbation will split a pair of degenerate
eigenfunctions $E$ of $H_{0}$ into eigenfunctions $B_{1}$ and $B_{2}$ of $H$
as follows from straightforward inspection of the character tables \ref
{tab:c4v} and \ref{tab:c2v}. Note that $(x,y)$ is basis for the irrep $E$ of
$C_{4v}$ and $x+y$ and $x-y$ are bases for $B_{1}$ and $B_{2}$,
respectively, of the modified $C_{2v}$. Besides, it is clear from $\sigma
_{v}(x+y)=-x+y$ and $\sigma _{v}^{\prime }(x+y)=x-y$ that $\hat{A}_{j}\psi
^{B_{1}}$ belongs to the irrep $B_{2}$; therefore $E_{n}^{B_{1}}=\left(
E_{m}^{B_{2}}\right) ^{*}$ as argued in section~\ref{sec:diagonalization}.

On the other hand, the perturbation corrections of first order for the pairs
of degenerate states $(A_{1},B_{1})$ and $(A_{2},B_{2})$ (coming from
dynamical symmetry) vanish as shown, for example, by $\left\langle \varphi
^{A_{1}}\right| H^{\prime }$ $\left| \varphi ^{A_{1}}\right\rangle
=\left\langle \varphi ^{B_{1}}\right| H^{\prime }$ $\left| \varphi
^{B_{1}}\right\rangle =\left\langle \varphi ^{A_{1}}\right| H^{\prime }$ $%
\left| \varphi ^{B_{1}}\right\rangle =0$. Consequently, the resulting
eigenfunctions of $H$ may have real eigenvalues for sufficiently small
values of $|g|$.

By means of projection operators\cite{T64,C90} we easily prove that the
connection between the eigenfunctions of $H_{0}$ and those of $H$ is given
by the following scheme:
\begin{eqnarray}
A_{1} &\rightarrow &A_{1}  \nonumber \\
A_{2} &\rightarrow &A_{2}  \nonumber \\
B_{1} &\rightarrow &A_{2}  \nonumber \\
B_{2} &\rightarrow &A_{1}  \nonumber \\
E &\rightarrow &B_{1},B_{2}  \label{eq:C4v->C2v}
\end{eqnarray}

As pointed out in section~\ref{sec:diagonalization}, in order to
obtain the eigenvalues of the models discussed in this paper we
resort to two independent methods: a collocation method\cite{AF10,
AFR11} and diagonalization of a truncated matrix representation
$\mathbf{H}$ of the Hamiltonian operator in a suitable basis set.
For the two-dimensional anharmonic oscillators discussed in this
section we choose the set of eigenfunctions of
$H_{HO}=p_{x}^{2}+p_{y}^{2}+x^{2}+y^{2}$. It is worth noting that
the coefficients of the characteristic polynomial
$|\mathbf{H}-E\mathbf{I}|=0$, where $\mathbf{I}$ is the identity
matrix, are real when we use the complete basis set, as discussed
by Fern\'{a}ndez\cite{F13}. On the other hand, if we resort to
symmetry-adapted basis sets $F^{B_{1}}$ and $F^{B_{2}}$ as
discussed in section~\ref{sec:symmetry}, then the coefficients of
the characteristic polynomials are complex\cite{FG14,FG14b}. Here
we diagonalize matrix representations $\mathbf{H}^{S}$ of the
Hamiltonian operator using symmetry-adapted basis functions for
the irreps $S=A_{1},A_{2},B_{1},B_{2}$ of the $C_{2v}$ point group
of Table~\ref{tab:c2v}.

The eigenvalues with eigenfunctions of symmetry $A_{1}$ and
$A_{2}$ are real for sufficiently small values of $g$. Pairs of
them approach each other and coalesce at exceptional points
$g_{c}$. For $g>g_{c}$ they become pairs of complex conjugate
numbers. On the other hand, the eigenvalues with eigenfunctions of
symmetry $B_{1}$ and $B_{2}$, which emerge from the irrep $E$ of
$C_{4v}$, appear to be complex for all $g>0$. This result, like
the one in reference \cite{FG14b}, also appears to contradict the
conjecture of Klaiman and Cederbaum\cite{KC08} outlined in
section~\ref{sec:symmetry}.

In the case of Hermitian operators there is the well known
non-crossing rule\cite{T37,RNBB72} that states that two
eigenvalues with eigenfunctions of the same symmetry do not cross
when they are plotted as functions of a parameter in the
Hamiltonian operator. In the case of non-Hermitian operators, on
the other hand, there is the coalescence rule that states that
only eigenvalues with eigenfunctions of the same symmetry
coalesce. This rule is clearly illustrated by the states with
symmetry $A_{1}$ and $A_{2}$ and is an obvious consequence of the
fact that we can group the states into different subspaces
according to their PGS.

We can easily construct other models based on the same $H_{0}$ that exhibit
broken ST symmetry for sufficiently small $|g|$. For example, $H^{\prime
}=xy^{3}$ is a linear combination of functions of symmetry $A_{2}$ ($%
xy(x^{2}-y^{2})$) and $B_{2}$ ($xy(x^{2}+y^{2})$) of the point
group $C_{4v}$ and is also invariant under parity ($C_{2}$ in this
case). In addition to it, $H$ exhibits the same antiunitary
symmetries $\hat{A}_{1}=T\sigma _{v}$ and $\hat{A}_{2}=T\sigma
_{v}^{\prime }$ discussed above. However, in this case $H^{\prime
}$ is invariant under the unitary operations $\{E,C_{2}\}$ of the
point group $C_{2}$ with irreps $\{A,B\}$ (see
Table~\ref{tab:c2}), where we have obviously chosen
$C_{2}:(x,y)\rightarrow (-x,-y)$. Because of the perturbation the
symmetry of the eigenfunctions changes in the following way:
$\{A_{1},A_{2},B_{1},B_{2}\}\rightarrow A$, $E\rightarrow B$. In
this case the perturbation splits pairs of degenerate
eigenfunctions of $H_{0}$ of symmetry $E$ into eigenfunctions of
$H$ that belong to the irrep $B$ and have complex conjugate
eigenvalues. The characteristic polynomial $\left|
\mathbf{H}^{B}-E\mathbf{I}\right| =0$ exhibits real coefficients
but complex roots. The eigenvalues with eigenfunctions of symmetry
$A$ are real for sufficiently small values of $g$ and pairs of
them coalesce at exceptional points as discussed above. On the
other hand, the eigenvalues with eigenfunctions of symmetry $B$
are complex for sufficiently small values of $g>0$. However, some
pairs of complex conjugate eigenvalues exhibit an interesting
behaviour. For example, the two complex eigenvalues that stem from
$E^{(0)}\approx 12.7$ become real at $g\approx 0.064096$,
separate, then approach each other and coalesce at $g\approx
1.08979$ becoming complex again for larger $g$. This surprising
behaviour was not observed in the earlier papers on ST-symmetric
Hamiltonians with complex eigenvalues\cite{FG14, FG14b}.

A slight modification of the perturbation leads to completely different
results. For example, $H^{\prime }=xy^{2}$ belongs to the irrep $E$ of the
point group $C_{4v}$ and $H$ results to be invariant under the unitary
transformations $\{E,\sigma \}$ of the point group $C_{s}$, where $\sigma
:(x,y)\rightarrow (x,-y)$. The irreps for $C_{s}$ are $A^{\prime }$ and $%
A^{\prime \prime }$ as shown in Table~\ref{tab:cs}. In this case $H$ is PT
symmetric, where $P=C_{2}$, and the perturbation connects the symmetry of
the eigenfunctions of $H_{0}$ and $H$ in the following way:
\begin{eqnarray}
A_{1} &\rightarrow &A^{\prime }  \nonumber \\
A_{2} &\rightarrow &A^{\prime \prime }  \nonumber \\
B_{1} &\rightarrow &A^{\prime }  \nonumber \\
B_{2} &\rightarrow &A^{\prime \prime }  \nonumber \\
E &\rightarrow &A^{\prime },A^{\prime \prime }.
\end{eqnarray}
Since the four matrix elements of $H^{\prime }$ between a pair of
$E$ eigenfunctions of $H_{0}$ vanish, then the perturbation
corrections of first order also vanish and the eigenvalues are
expected to be real for $0\leq g<g_{c}$. Numerical results confirm
our argument based on point-group symmetry and perturbation
theory: all the eigenvalues are real for sufficiently small values
of $g$. As $g$ increases pairs of eigenvalues coalesce at
exceptional points as expected; however some of them exhibit an
interesting behaviour. For example, one of the $A^{\prime }$
eigenvalues stemming from $E^{(0)}\approx 27.59$ and one stemming
from $E^{(0)}\approx 27.91$ approach each other and coalesce. They
become a pair of complex conjugate numbers for some values of $g$
and then separate again as real eigenvalues. One of the
resulting branches and the other real eigenvalue stemming from $%
E^{(0)}\approx 27.59$ coalesce at another exceptional point. On the other
hand, the other branch and an eigenvalue stemming from $E^{(0)}\approx 30.33$
coalesce at another exceptional point.

We can also build a non-Hermitian oscillator with unbroken ST symmetry by
reducing the geometrical symmetry of $H_{0}$. If we choose $%
H_{0}=p_{x}^{2}+p_{y}^{2}+\alpha _{x}x^{4}+\alpha _{y}y^{4}$, with $\alpha
_{x}\neq \alpha _{y}$, the point group for $H_{0}$ is $C_{2v}$ with only
one-dimensional irreps. Let us consider, for example, the perturbation $%
H^{\prime }=xy$ that is invariant under parity ($C_{2}$). In this case $H$
is invariant under the antiunitary transformations $\hat{A}_{1}=T\sigma _{v}$
and $\hat{A}_{2}=T\sigma _{v}^{\prime }$ already introduced above and,
therefore, ST symmetric. However, in this case all the perturbation
corrections of first order vanish and numerical calculations suggest that
the eigenvalues of this Hamiltonian are real for all $0\leq g<g_{c}$\cite
{KC08}.

We can construct other interesting models by enclosing oscillators in boxes
with impenetrable walls and suitable geometries. For example,
\begin{equation}
H_{0}=p_{x}^{2}+p_{y}^{2},  \label{eq:H0_box_2D}
\end{equation}
with the boundary conditions $\psi (\pm 1,y)=0$ and $\psi (x,\pm
1)=0$ (square box of length $L=2$). In this case we can also
choose $C_{4v}$ to describe the symmetry of the Hermitian part.
When $H^{\prime }=xy^{2}$ the eigenvalues are real for all $0\leq
g<g_{c}$, while $H^{\prime }=xy$ produces complex eigenvalues of
symmetry $B_{1}$ and $B_{2}$ for sufficiently small $g>0$. These
two models have already been discussed by Fern\'{a}ndez and
Garcia\cite{FG14b}. On the other hand, $H^{\prime }=xy^{3}$ leads
to complex conjugate eigenvalues of symmetry $B$ for small $g>0$
but some pairs of them separate into real ones, then approach each
other and coalesce again at exceptional points. Since the symmetry
of the Hermitian and non-Hermitian parts is identical to the
examples discussed above the behaviour of the eigenvalues for the
box models and the anharmonic oscillators is quite similar. The
main difference is that in the case of the box models the
exceptional points appear at much larger values of $g$.

The two dimensional isotropic harmonic oscillator
\begin{equation}
H_{0}=p_{x}^{2}+p_{y}^{2}+x^{2}+y^{2},  \label{eq:H0_x2_y2}
\end{equation}
is invariant under the two-dimensional rotation group (we can choose the $%
C_{\infty v}$ point group\cite{T64,C90}). In this case we draw the
same conclusions as before. When $H^{\prime }=xy^{2}$ we have the
non-Hermitian version of the Barbanis Hamiltonian that has been
widely studied\cite {BDMS01,NA02,N05,BTZ06,W09,BW12,HV13}.
Numerical calculations based on the diagonalization method,
perturbation theory and other approaches suggest that its
eigenvalues are real for all $0\leq g<g_{c}$, where $g_{c}$ is the
exceptional point closest to the origin. If, on the other hand,
$H^{\prime }=xy$ then some of the eigenvalues of the resulting
exactly-solvable model are complex for all $g$\cite{FG14}.

The models discussed in this section clearly show that ST symmetry does not
guarantee a real spectrum unless $S=P$. Note that of all the perturbations
studied above only $H^{\prime }=xy^{2}$ satisfies this condition.

\section{Three-dimensional models}

\label{sec:3D_models}

We first consider the Hermitian Hamiltonian oscillator
\begin{equation}
H_{0}=p_{x}^{2}+p_{y}^{2}+p_{z}^{2}+\alpha _{x}x^{4}+\alpha _{y}y^{4}+\alpha
_{z}z^{4},  \label{eq:H0_x4_y4_z4}
\end{equation}
where $\alpha _{x}$, $\alpha _{y}$ and $\alpha _{z}$ are real and positive.
If the three potential parameters $\alpha _{q}$ are different then this
operator is invariant under the unitary transformations of the point group $%
C_{i}$. Since its eigenfunctions belong to the one-dimensional irreps $A_{g}$
and $B_{g}$, one expects the eigenvalues of any space-time symmetric
Hamiltonian $H$ built from it to have real eigenvalues for some interval of
parameter values $0\leq g<g_{c}$. If, for example, $\alpha _{x}=\alpha
_{y}\neq \alpha _{z}$ then $H_{0}$ is invariant under the operations of the
point group $C_{4v}$ and we expect results similar to those discussed in
Section~\ref{sec:2D_models}; that is to say: for some non-Hermitian
perturbations the eigenvalues may be complex for sufficiently small $g>0$.

Therefore, the most interesting case seems to be $\alpha _{x}=\alpha
_{y}=\alpha _{z}=\alpha $ and without loss of generality in what follows we
choose $\alpha =1$. In such a case $H_{0}$ is invariant under the unitary
transformations of the point group $O_{h}$ shown in Table~\ref{tab:oh}. The
degeneracy of the energy levels of a quantum-mechanical model with this PGS
has been recently discussed\cite{F13b,HCL13}.

If $\{i,j,k\}_{P}$ denotes all distinct permutations of the subscripts in
the eigenfunctions of $H_{0}$ $\varphi _{i\,j\,k}(x,y,z)=\phi _{i}(x)\phi
_{j}(y)\phi _{k}(z)$, $i,j,k=0,1,\ldots $, then their symmetry and dimension
of the eigenspaces are given by\ (see reference \cite{F13b} for a discussion
of another quantum-mechanical problem with the same PGS):

\begin{equation}
\begin{array}{ll}
\{2n,2n,2n\} & A_{1g} \\
\{2n+1,2n+1,2n+1\} & A_{2u} \\
\{2n+1,2n+1,2m\}_{P} & T_{2g} \\
\{2n,2n,2m+1\}_{P} & T_{1u} \\
\{2n,2n,2m\}_{P} & A_{1g},E_{g} \\
\{2n+1,2n+1,2m+1\}_{P} & A_{2u},E_{u} \\
\{2n,2m,2k\}_{P} & A_{1g},A_{2g},E_{g},E_{g} \\
\{2n+1,2m+1,2k+1\}_{P} & A_{1u},A_{2u},E_{u},E_{u} \\
\{2n,2m,2k+1\}_{P} & T_{1u},T_{2u} \\
\{2n+1,2m+1,2k\}_{P} & T_{1g},T_{2g}
\end{array}
.  \label{eq:degeneracy_2}
\end{equation}
The dynamical symmetries that are responsible for the degeneracy of
eigenfunctions belonging to different irreps (which cannot be explained by
PGS) are given by the Hermitian operators
\begin{eqnarray}
O_{1} &=&2p_{x}^{2}+2x^{4}-p_{y}^{2}-y^{4}-p_{z}^{2}-z^{4}  \nonumber \\
O_{2} &=&2p_{y}^{2}+2y^{4}-p_{x}^{2}-x^{4}-p_{z}^{2}-z^{4},
\end{eqnarray}
which belong to the irrep $E_{g}$. In order to obtain them we simply apply
the projection operator $P^{E_{g}}$ to the two pairs of functions $%
(x^{2},y^{2})$ and $(x^{4},y^{4})$ as discussed elsewhere\cite{F13b}.

If we take into account that $T_{1g}\otimes T_{1g}=T_{2g}\otimes
T_{2g}=T_{1u}\otimes T_{1u}=T_{2u}\otimes T_{2u}=A_{1g}\oplus
E_{g}\oplus T_{1g}\oplus T_{2g}$, then we realize that a
perturbation $H^{\prime }$ belonging to the irrep $T_{2g}$ will
split those degenerate energy levels and produce complex
eigenvalues for sufficiently small $g>0$. According to the
character table in Table~\ref{tab:oh}, any linear combination of
the functions $xy $, $xz$ and $yz$ will suffice. If, for example,
we choose $H^{\prime }=z(x+y) $, then the Hamiltonian $H$ is
invariant under the antiunitary
transformations $\hat{A}_{1}=C_{2}^{\prime }T$ and $\hat{A}_{2}=\sigma _{h}T$%
, where $C_{2}^{\prime }:(x,y,z)\rightarrow (-x,-y,z)$ and $\sigma
_{h}:(x,y,z)\rightarrow (x,y,-z)$. The resulting space-time invariant
Hamiltonian $H$ is also invariant under the unitary transformations of the
point group $C_{2h}$ if we choose them in the following way: $%
C_{2}:(x,y,z)\rightarrow (-y,-x,-z)$, $\hat{\imath}:(x,y,z)\rightarrow
(-x,-y,-z)$ and $\sigma _{h}:(x,y,z)\rightarrow (y,x,z)$ as shown in Table~%
\ref{tab:c2h}. Note that $H^{\prime }$ is invariant under parity inversion $%
P=\hat{\imath}$.

The connection between the eigenfunctions of $H_{0}$ and $H$ is given by
\begin{eqnarray}
A_{1g} &\rightarrow &A_{g}  \nonumber \\
A_{2g} &\rightarrow &B_{g}  \nonumber \\
E_{g} &\rightarrow &A_{g},B_{g}  \nonumber \\
T_{1g} &\rightarrow &A_{g},2B_{g}  \nonumber \\
T_{2g} &\rightarrow &2A_{g},B_{g}  \nonumber \\
A_{1u} &\rightarrow &A_{u}  \nonumber \\
A_{2u} &\rightarrow &B_{u}  \nonumber \\
E_{u} &\rightarrow &A_{u},B_{u}  \nonumber \\
T_{1u} &\rightarrow &A_{u},2B_{u}  \nonumber \\
T_{2u} &\rightarrow &2A_{u},B_{u},  \label{eq:Oh->C2h}
\end{eqnarray}
and those corresponding to the three-dimensional irreps will produce complex
eigenvalues for $g>0$ as argued above. Equations (\ref{eq:degeneracy_2}) and
(\ref{eq:Oh->C2h}) together summarize the splitting of the energy levels of
an $O_{h}$ Hermitian Hamiltonian by a $C_{2h}$ non-Hermitian perturbation.

Table~\ref{tab:E0x4y4z4} shows the lowest eigenvalues of $H_{0}$
calculated by means of the Riccati-Pad\'{e} method\cite{FMT89a,
FMT89b} and the quantum numbers of their corresponding states. The
eigenvalue stemming from $E^{(0)}\approx 3.18$ of symmetry $A_{g}$
is real for all $g$. The next one starting at $E^{(0)}\approx
5.92$ splits into one real $A_{u}$ and two complex $B_{u}$. The
next one at $E^{(0)}\approx 8.66$ gives rise to one real
eigenvalue $B_{g}$ and two complex ones $A_{g}$. The next one at
$E^{(0)}\approx 9.58$ leads to three real eigenvalues: two $A_{g}$
and one $B_{g}$. The two real eigenvalues $B_{g}$ approach each
other and coalesce at an exceptional point $g_{c}\approx 1.0713$
where they become a pair of complex conjugate numbers. The next
eigenvalue at $E^{(0)}\approx 11.40$ is real and $B_{u}$. The
sixth-dimensional eigenspace for $E^{(0)}\approx 12.32$ consists
of three functions $T_{1u}$ and three $T_{2u}$. The former split
into two complex eigenvalues $A_{u}$ and one real $B_{u}$. The
latter split into two complex $B_{u}$ and one real $A_{u}$. The
two real eigenvalues $B_{u}$ just
mentioned approach each other and coalesce at an exceptional point $%
g_{c}\approx 1.3064$. The eigenfunctions of symmetry $T_{1u}$ with
eigenvalue $E^{(0)}\approx 13.77$ are most interesting. They split into two
complex $B_{u}$ and one real $A_{u} $; however the two complex $B_{u}$
eigenvalues become real at $g\approx 0.018578$, separate and then approach
each other to coalesce at an exceptional point $g_{c}\approx 0.83161$. We
have already encountered this behaviour in one of the two-dimensional
examples discussed in section~\ref{sec:2D_models}.

Another model with the same symmetry is given by
\begin{equation}
H_{0}=p_{x}^{2}+p_{y}^{2}+p_{z}^{2},  \label{eq:H0_box_3D}
\end{equation}
with the boundary conditions $\psi (\pm 1,y,z)=\psi (x,\pm 1,z)=\psi
(x,y,\pm 1)=0$. The point group for this system is also $O_{h}$ and was
discussed in detail by Fern\'{a}ndez\cite{F13b} and Hern\'{a}ndez-Castillo
and Lemus\cite{HCL13}. The dimensionless eigenvalues and eigenfunctions are
\begin{eqnarray}
E_{n_{1}n_{2}n_{3}} &=&\frac{\pi ^{2}}{4}(n_{1}^{2}+n_{2}^{2}+n_{3}^{2})
\nonumber \\
\psi _{n_{1}n_{2}n_{3}}(x,y,z) &=&\sin \left[ \frac{n_{1}\pi (x+1)}{2}%
\right] \sin \left[ \frac{n_{2}\pi (y+1)}{2}\right] \sin \left[ \frac{%
n_{3}\pi (z+1)}{2}\right] ,  \nonumber \\
&&  \label{eq:box3D_eigenval_eigenfun}
\end{eqnarray}
where $n_{1},n_{2},n_{3}=1,2,\ldots $. The symmetry of the eigenfunctions is
similar to the scheme in equation (\ref{eq:degeneracy_2}) by substituting $%
(2n_{1}-1,2n_{2}-1,2n_{3}-1)$ for $(2m,2n,2k)$ and $(2n_{1},2n_{2},2n_{3})$
for $(2m+1,2n+1,2k+1)$ \cite{F13b}.

Obviously, the same parity-invariant non-Hermitian perturbations discussed
above lead to complex eigenvalues for $g\neq 0$. However, in this case we
can easily calculate the perturbation corrections of first order
analytically and show which eigenvalues are complex when $g\neq 0$. For
example, for $H^{\prime }=z(x+y)$ we easily obtain the following
perturbation expansions for the eigenvalues:
\begin{eqnarray}
\{1,1,1\} &\rightarrow &\frac{3\pi ^{2}}{4}+O(\lambda ^{2})  \nonumber \\
\{1,1,2\}_{p} &\rightarrow &\left\{
\begin{array}{l}
\frac{3\pi ^{2}}{2}-\frac{1024\sqrt{2}\lambda }{81\pi ^{4}}+O(\lambda ^{2})
\\
\frac{3\pi ^{2}}{2}+O(\lambda ^{2}) \\
\frac{3\pi ^{2}}{2}+\frac{1024\sqrt{2}\lambda }{81\pi ^{4}}+O(\lambda ^{2})
\end{array}
\right.  \nonumber \\
\{1,2,2\}_{p} &\rightarrow &\left\{
\begin{array}{l}
\frac{9\pi ^{2}}{4}-\frac{1024\sqrt{2}\lambda }{81\pi ^{4}}+O(\lambda ^{2})
\\
\frac{9\pi ^{2}}{4}+O(\lambda ^{2}) \\
\frac{9\pi ^{2}}{4}+\frac{1024\sqrt{2}\lambda }{81\pi ^{4}}+O(\lambda ^{2})
\end{array}
\right.  \nonumber \\
\{1,1,3\}_{p} &\rightarrow &\left\{
\begin{array}{l}
\frac{11\pi ^{2}}{4}+O(\lambda ^{2}) \\
\frac{11\pi ^{2}}{4}+O(\lambda ^{2}) \\
\frac{11\pi ^{2}}{4}+O(\lambda ^{2})
\end{array}
\right.  \nonumber \\
\{2,2,2\} &\rightarrow &3\pi ^{2}+O(\lambda ^{2})  \nonumber \\
\{1,2,3\}_{p} &\rightarrow &\left\{
\begin{array}{l}
\frac{7\pi ^{2}}{2}-\frac{1024\sqrt{922066}\lambda }{50625\pi ^{4}}%
+O(\lambda ^{2}) \\
\frac{7\pi ^{2}}{2}-\frac{1024\sqrt{922066}\lambda }{50625\pi ^{4}}%
+O(\lambda ^{2}) \\
\frac{7\pi ^{2}}{2}+O(\lambda ^{2}) \\
\frac{7\pi ^{2}}{2}+O(\lambda ^{2}) \\
\frac{7\pi ^{2}}{2}+\frac{1024\sqrt{922066}\lambda }{50625\pi ^{4}}%
+O(\lambda ^{2}) \\
\frac{7\pi ^{2}}{2}+\frac{1024\sqrt{922066}\lambda }{50625\pi ^{4}}%
+O(\lambda ^{2})
\end{array}
\right.  \nonumber \\
\{2,2,3\}_{p} &\rightarrow &\left\{
\begin{array}{l}
\frac{17\pi ^{2}}{4}-\frac{9216\sqrt{2}}{625\pi ^{4}}+O(\lambda ^{2}) \\
\frac{17\pi ^{2}}{4}+O(\lambda ^{2}) \\
\frac{17\pi ^{2}}{4}+\frac{9216\sqrt{2}}{625\pi ^{4}}+O(\lambda ^{2})
\end{array}
\right. ,
\end{eqnarray}
for the first eigenvalues. Those states with nonzero perturbation correction
of first order are expected to be complex for sufficiently small $|g|$. The
splitting of the energy levels of $H_{0}$ by the perturbation $H^{\prime }$
is also summarized by equations (\ref{eq:degeneracy_2}) and (\ref{eq:Oh->C2h}%
) with the substitutions already mentioned above. For example, the three
eigenfunctions of order zero generated by the label permutations $%
\{1,1,2\}_{P}$ are basis for the irrep $T_{1u}$ when $g=0$ and split into
two $B_{u}$ with complex conjugate eigenvalues and one $A_{u}$ with real
eigenvalue.

\section{Conclusions}
\label{sec:conclusions}

Throughout this paper we have discussed non-Hermitian Hamiltonian
operators of the form (\ref{eq:H_gen}) where the Hermitian and
non-Hermitian parts exhibit several different PGS. In each case we
have clearly indicated how the energy levels of $H_{0}$ behave
when the perturbation is turned on. The nature of the resulting
eigenvalues of $H$ depend on the symmetry of both $H_{0}$ and
$H^{\prime }$. PGS and perturbation theory enable us to predict
whether there is a chance that the eigenvalues of $H$ are real for
some values of the strength parameter $g$. If the perturbation
correction of first order is nonzero for at least one state then
we expect complex eigenvalues for sufficiently small $|g|$.
Complex eigenvalues may become real for some values of $g$ but it
is unlikely that such intervals overlap to produce an island of
real eigenvalues for all the states of the model. It is worth
noting that space-time symmetry only tells us that the eigenvalues
of the non-Hermitian Hamiltonian are either real or appear in
pairs of complex conjugate numbers. On the other hand, the
analysis based on perturbation theory provides a much clearer
indication of whether there is any chance that the eigenvalues are
real for sufficiently small nonzero values of $g$.

One of the main conclusions of this paper is that ST symmetry is not a
satisfactory generalization of PT symmetry, except when the full point group
of symmetry for $H_{0}$ is Abelian. An ST-symmetric Hamiltonian may exhibit
complex eigenvalues for sufficiently small $|g|$ when the unitary operation $%
S$ is different from the parity inversion $P$. On the other hand, PT
symmetry has led to real eigenvalues for all $0<g<g_{c}$ in all the cases
studied so far.

\section*{Acknowledgements}
This report has been financially supported by PIP No.
11420110100062 (Consejo Nacional de Investigaciones Cientificas y
Tecnicas, Rep\'{u}blica Argentina)

\begin{table}[]
\caption{Character table for $C_{4v}$ point group}
\label{tab:c4v}
\begin{tabular}{l|rrrrr|l|l}
$C_{4v}$ & $E$ & $2C_4$ & $C_2$ & $2\sigma_v$ & $2\sigma_d$ &  &  \\ \hline
$A_1$ & 1 & 1 & 1 & 1 & 1 & $z$ & $x^2+y^2,z^2$ \\
$A_2$ & 1 & 1 & 1 & -1 & -1 & $R_z$ &  \\
$B_1$ & 1 & -1 & 1 & 1 & -1 &  & $x^2-y^2$ \\
$B_2$ & 1 & -1 & 1 & -1 & 1 &  & $xy$ \\
$E$ & 2 & 0 & -2 & 0 & 0 & $(x,y)(R_x, R_y)$ & $(xz, yz)$%
\end{tabular}
\end{table}

\begin{table}[]
\caption{Character table for the modified $C_{2v}$ point group}
\label{tab:c2v}
\begin{tabular}{l|rrrr|l|l}
$C_{2v}$ & $E$ & $C_2$ & $\sigma_d$ & $\sigma_d^{\prime}$ &  &  \\ \hline
$A_1$ & 1 & 1 & 1 & 1 &  & $x^2+y^2,xy$ \\
$A_2$ & 1 & 1 & -1 & -1 &  & $x^2-y^2$ \\
$B_1$ & 1 & -1 & 1 & -1 & $x+y$ &  \\
$B_2$ & 1 & -1 & -1 & 1 & $x-y$ &
\end{tabular}
\end{table}

\begin{table}[]
\caption{First eigenvalues of $H_0$ (\ref{eq:H0_x4_y4})}
\label{tab:E0x4y4}%
\begin{tabular}{D{.}{.}{20}cc}

 \multicolumn{1}{c}{$E_{n_1 n_2}$}   & $n_1$ & $n_2$  \\

 2.1207241809683657991&  0&   0 \\
 4.8600351202855770683&  0&   1 \\
 4.8600351202855770683&  1&   0 \\
 7.5993460596027883375&  1&   1 \\
 8.5160600284709212917&  0&   2 \\
 8.5160600284709212917&  2&   0 \\
 11.255370967788132561&  1&   2 \\
 11.255370967788132561&  2&   1 \\
 12.70510760186234492 &  0&   3 \\
 12.70510760186234492 &  3&   0 \\
 14.911395875973476784&  2&   2 \\
 15.444418541179556189&  1&   3 \\
 15.444418541179556189&  3&   1 \\
 17.322188109334408837&  0&   4 \\
 17.322188109334408837&  4&   0 \\
 19.100443449364900413&  2&   3 \\
 19.100443449364900413&  3&   2 \\

\end{tabular}
\end{table}

\begin{table}[]
\caption{Character table for $C_{2}$ point group}
\label{tab:c2}
\begin{tabular}{l|rr|l|l}
$C_{2}$ & $E$ & $C_2$ &  &  \\ \hline
$A$ & 1 & 1 &  & $x^2,y^2,xy$ \\
$B$ & 1 & -1 & $x,y$ &
\end{tabular}
\end{table}

\begin{table}[]
\caption{Character table for $C_{s}$ point group}
\label{tab:cs}
\begin{tabular}{l|rr|l|l}
$C_{s}$ & $E$ & $\sigma$ &  &  \\ \hline
$A^{\prime}$ & 1 & 1 & $x$ & $x^2,y^2$ \\
$A^{\prime\prime}$ & 1 & -1 & $y$ & $xy$%
\end{tabular}
\end{table}

\begin{table}[]
\caption{Character table for $O_h$ point group}
\label{tab:oh}{\tiny
\begin{tabular}{l|rrrrrrrrrr|l|l}
$O_h$ & $E$ & $8C_3$ & $6C_2$ & $6C_4$ & $3C_2(=C_4^2)$ & $i$ & $6S_4$ & $%
8S_6$ & $3\sigma_h$ & $6\sigma_d$ &  &  \\ \hline
A1g & 1 & 1 & 1 & 1 & 1 & 1 & 1 & 1 & 1 & 1 &  & $x^2+y^2+z^2$ \\
A2g & 1 & 1 & -1 & -1 & 1 & 1 & -1 & 1 & 1 & -1 &  &  \\
Eg & 2 & -1 & 0 & 0 & 2 & 2 & 0 & -1 & 2 & 0 &  & $(2z^2-x^2-y^2, x^2-y^2)$
\\
T1g & 3 & 0 & -1 & 1 & -1 & 3 & 1 & 0 & -1 & -1 & $(R_x, R_y, R_z)$ &  \\
T2g & 3 & 0 & 1 & -1 & -1 & 3 & -1 & 0 & -1 & 1 & $(xz, yz, xy)$ &  \\
A1u & 1 & 1 & 1 & 1 & 1 & -1 & -1 & -1 & -1 & -1 &  &  \\
A2u & 1 & 1 & -1 & -1 & 1 & -1 & 1 & -1 & -1 & 1 &  &  \\
Eu & 2 & -1 & 0 & 0 & 2 & -2 & 0 & 1 & -2 & 0 &  &  \\
T1u & 3 & 0 & -1 & 1 & -1 & -3 & -1 & 0 & 1 & 1 & $(x, y, z)$ &  \\
T2u & 3 & 0 & 1 & -1 & -1 & -3 & 1 & 0 & 1 & -1 &  &
\end{tabular}
}
\end{table}

\begin{table}[]
\caption{Character table for $C_{2h}$ point group}
\label{tab:c2h}
\begin{tabular}{l|rrrr|l|l}
$C_{2h}$ & $E$ & $C_2$ & $i$ & $\sigma_h$ &  &  \\ \hline
$A_g$ & 1 & 1 & 1 & 1 &  & $x^2+y^2,z(x+y),xy,z^2$ \\
$B_g$ & 1 & -1 & 1 & -1 &  & $x^2-y^2,z(x-y)$ \\
$A_u$ & 1 & 1 & -1 & -1 & $x-y$ &  \\
$B_u$ & 1 & -1 & -1 & 1 & $x+y,z$ &
\end{tabular}
\end{table}

\begin{table}[]
\caption{First eigenvalues of $H_0$ (\ref{eq:H0_x4_y4_z4}) with $%
\alpha_x=\alpha_y=\alpha_z=1$.}
\label{tab:E0x4y4z4}%
\begin{tabular}{D{.}{.}{20}ccc}

 \multicolumn{1}{c}{$E_{n_1 n_2 n_3}$}   & $n_1$ & $n_2$ & $n_3$ \\
 3.1810862714525486987 & 0 &  0 &  0 \\
 5.9203972107697599679 & 0 &  0 &  1 \\
 5.9203972107697599679 & 0 &  1 &  0 \\
 5.9203972107697599679 & 1 &  0 &  0 \\
 8.6597081500869712372 & 0 &  1 &  1 \\
 8.6597081500869712372 & 1 &  0 &  1 \\
 8.6597081500869712372 & 1 &  1 &  0 \\
 9.5764221189551041913 & 0 &  0 &  2 \\
 9.5764221189551041913 & 0 &  2 &  0 \\
 9.5764221189551041914 & 2 &  0 &  0 \\
 11.399019089404182506 & 1 &  1 &  1 \\
 12.31573305827231546  & 1 &  0 &  2 \\
 12.31573305827231546  & 1 &  2 &  0 \\
 12.31573305827231546  & 2 &  0 &  1 \\
 12.31573305827231546  & 2 &  1 &  0 \\
 12.31573305827231546  & 0 &  1 &  2 \\
 12.31573305827231546  & 0 &  2 &  1 \\
 13.76546969234652782  & 0 &  0 &  3 \\
 13.76546969234652782  & 0 &  3 &  0 \\

\end{tabular}
\end{table}

\end{document}